\documentclass[preprint,12pt]{elsarticle}

\usepackage{graphicx}
\usepackage{hyperref}
\usepackage{amssymb}

\usepackage{lineno}

\journal{ArXiv}

\begin{document}

\begin{frontmatter}


\title{BioSpaun: A large-scale behaving brain model with complex neurons}



\author{Chris Eliasmith}
\author{Jan Gosmann}
\author{Xuan Choo}

\address{Centre for Theoretical Neuroscience, University of Waterloo, Waterloo, Ontario, Canada}

\begin{abstract}
We describe a large-scale functional brain model that includes detailed, conductance-based, compartmental models of individual neurons.  We call the model BioSpaun, to indicate the increased biological plausibility of these neurons, and because it is a direct extension of the Spaun model \cite{Eliasmith2012b}.  We demonstrate that including these detailed compartmental models does not adversely affect performance across a variety of tasks, including digit recognition, serial working memory, and counting.  We then explore the effects of applying TTX, a sodium channel blocking drug, to the model.  We characterize the behavioral changes that result from this molecular level intervention.  We believe this is the first demonstration of a large-scale brain model that clearly links low-level molecular interventions and high-level behavior.
\end{abstract}


\end{frontmatter}


\section{Introduction}
\label{S:intro}

Recently, several large-scale brain models have been described.  These include a biophysically detailed model from Markram's group in the Human Brain Project (HBP) \cite{Markram2015}, which includes about 31,000 compartmental neurons and 37 million synapses, modelled with many equations per cell.  This model is large-scale because of the amount of computation required to simulate its behavior at this level of biological detail.  Another model reported earlier by the Synapse project has simulated 500 billion neurons -- more than 5x the number in the human brain -- although each neuron is much simpler than those in the HBP model, and the connectivity is far more limited \cite{modha2012, Merolla2014a}.  We have also previously proposed a large-scale model that includes 2.5 million neurons, 8 billion connections, and, unlike these other large-scale models, exhibits a wide variety of cognitive behavior \cite{Eliasmith2012b}. However, our model uses simple leaky integrate-and-fire neurons, and for this reason Markram has claimed ``It's not a brain model" \cite{Sanders2013}.

In this paper we incorporate detailed compartmental models of the type used in the recent HBP model into different cortical areas of our large-scale, behaving brain model. We refer to this augmented model as ``BioSpaun''.  We show that the behavior of the original Spaun model is not adversely affected by changing the neuron model.  We further show that the additional complexity can be used to test hypotheses not possible with the original model. Specifically, we demonstrate that BioSpaun can be used to simulate the effects of adding the drug tetrodotoxin (TTX) to these areas of cortex.  We perform this manipulation to both visual cortex and frontal cortex, demonstrating performance declines related to the dosage of drug applied, both within and across different tasks.  While much remains to be done to verify the accuracy of these simulations \emph{in vivo}, we believe this is the first demonstration of a large-scale behaving neural model that includes a high degree of biophysical detail.  Integrating these two aspects of brain modeling provides a new method for testing low-level molecular and other physiological interventions on high-level behavior.

\section{Methods}
\subsection{Modeling approach}
The Neural Engineering Framework (NEF) identifies three quantitatively specified principles that can be used to implement nonlinear dynamical systems in a spiking neural substrate \cite{Eliasmith2003m}. These methods have been used to propose novel models of a wide variety of neural systems including parts of the rodent navigation system \cite{Conklin2005b}, tactile working memory in monkeys \cite{Singh2006b}, and simple decision making in humans \cite{Litt2008u} and rats \cite{bekolay2014a}. These methods have also been used to better understand more general issues about neural function, such as how the variability of neural spike trains and the timing of individual spikes relates to information that can be extracted from spike patterns \cite{Tripp2007c}, and how mixed weight neuron models can be transformed into models respecting Dale’s Principle, while preserving function \cite{Parisien2008c}.  

Conceptually, the NEF can be thought of as a “neural compiler” which allows a researcher to specify a computation as a general nonlinear dynamical system in some state space, which is then implemented in a spiking neural substrate using an efficient  optimization method.  There are several sources for detailed descriptions of these methods \cite{Eliasmith2003m, Eliasmith2005p, Stewart2014}, so we do not describe them here. Centrally, the NEF answers questions about \emph{how} neural systems might compute, but it does not address the issue of 
\emph{what}, specifically, is computed by biological brains.  

In our more recent work, we address this second question by proposing a general neural architecture that includes specific functional hypotheses.  We call this proposal the Semantic Pointer Architecture (SPA; \cite{Eliasmith2013}).  The SPA identifies a generic means of characterizing neural representation, “semantic pointers,” that are used to capture central features of perceptual, motor, and cognitive representation. The SPA uses semantic pointers to address perceptual categorization, motor control, working and long-term memory, as well as conceptual binding and structure representation.  As well, the SPA includes a characterization of cognitive control that relies on basal ganglia and thalamic interactions with cortex for understanding action selection.

We have developed several models using the SPA that address a variety of cognitive abilities. For instance, we have demonstrated the encoding and decoding of the 114,000 concepts and their relations in the WordNet database \cite{crawford2015}.  We have shown an SPA model that matches human performance on the full Raven’s Progressive Matrices (RPM) intelligence test, and demonstrated its ability to capture aging effects through biological manipulation \cite{Rasmussen201}.  We have described how the SPA naturally unifies the three most prevalent theories of conceptual representation\cite{blouw2015}.  We have also shown simple models of language parsing \cite{Stewart2014a}, instruction following \cite{choo2013}, and the n-back task \cite{gosmann2015}.  Together, we believe this body of works demonstrates a uniquely scalable and biologically plausible approach to understanding cognitive function.

\subsection{The Semantic Pointer Architecture Unified Network (Spaun)}
\label{S:spaun}

\begin{figure}
  \centering\includegraphics[width=\columnwidth]{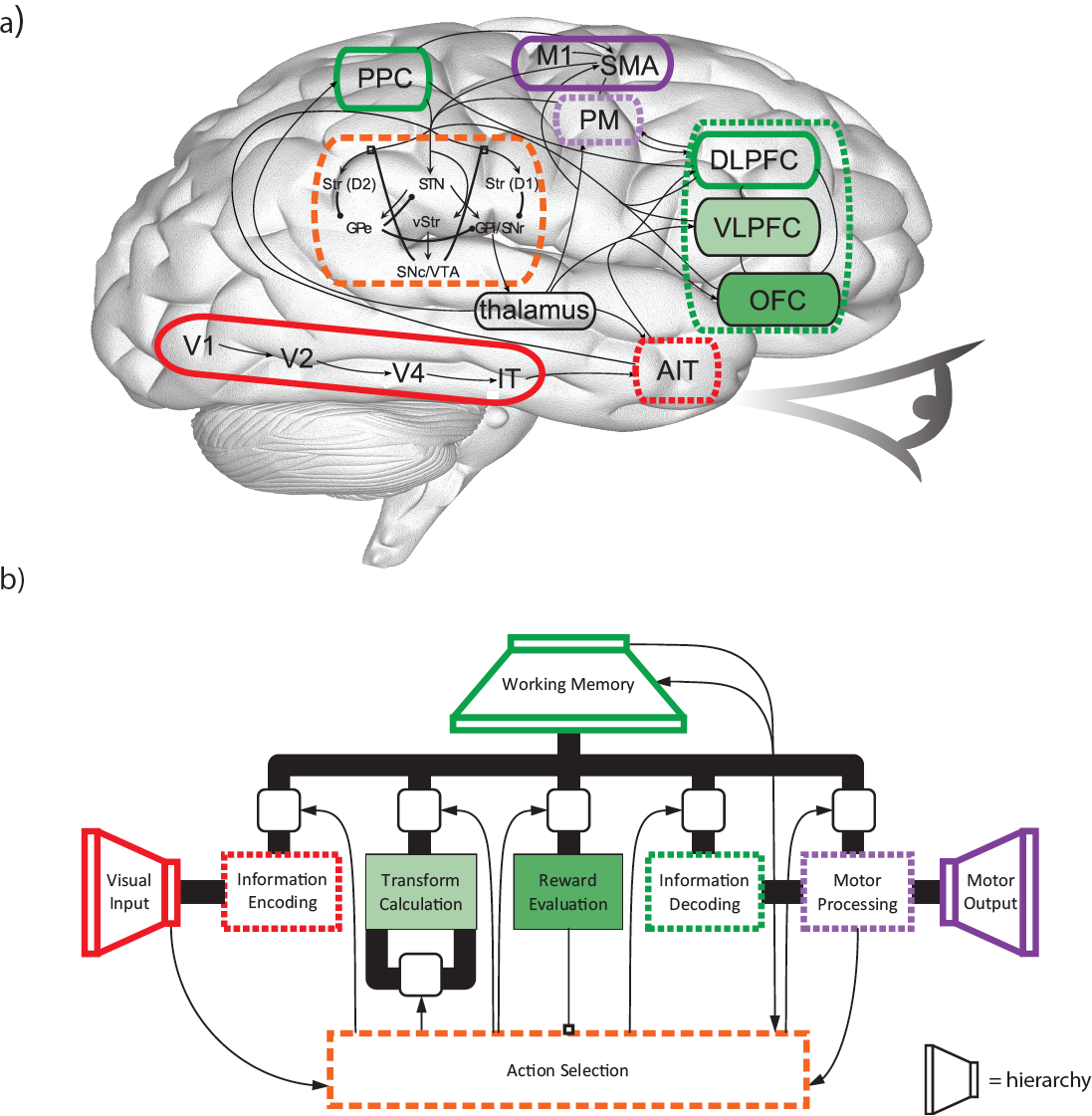}
  \caption{Functional and anatomical architecture of Spaun.  (a) Neuroanatomical architecture of Spaun, with matching colors and line styles indicating corresponding components in the functional architecture in b. Abbreviations: V1/V2/V4 (primary/secondary/extrastriate visual cortex), AIT/IT (anterior/inferotemporal cortex), DLPFC/VLPFC/OFC (dorso-lateral/ventro-lateral/orbito- frontal cortex), PPC (posterior parietal cortex), M1 (primary motor cortex), SMA (supplementary motor area), PM (premotor cortex), v/Str (ventral/striatum), STN (subthalamic nucleus), GPe/i (globus pallidus externus/internus), SNc/r (substantia nigra pars compacta/reticulata), VTA (ventral tegmental area). (b) Functional architecture of Spaun.  The working memory, visual input, and motor output components represent hierarchies that compress/decompress neural representations between different representational spaces.  The action selection component chooses which action to execute given the current state of the rest of the system.  The five internal subsystems, from left to right, are used to 1) map visual inputs to conceptual representations, 2) induce relationships between representations, 3) associate input with reward, 4) map conceptual representations to motor actions, and 5) map motor actions to specific patterns of movement. Reproduced from \cite{Eliasmith2012b}.}
  \label{fig-spaun}
\end{figure}

To demonstrate the SPA in detail, we proposed a mechanistic, functional model of the brain that uses 2.5 million spiking neurons, has about 8 billion synaptic connections, and performs 8 different tasks \cite{Eliasmith2012b,Eliasmith2013}.  We refer to this large-scale neural model as the Semantic Pointer Architecture Unified Network (Spaun).  Spaun consists of leaky integrate-and-fire (LIF) model neurons.  The physiological and tuning properties of the cells are statistically matched to the various anatomical areas included in the model.  There are about 20 anatomical areas accounted for (see Figure \ref{fig-spaun}).  Four types of neurotransmitters are included in the model (GABA, AMPA, NMDA, and Dopamine), and their known time constants and synaptic effects are simulated.  

What makes Spaun unique among large-scale brain models is its functional abilities. Spaun receives input from the environment through its single eye, which is shown images of handwritten or typed digits and letters, and it manipulates the environment by moving a physically modelled arm, which has mass, length, inertia, and so on.  Spaun uses these natural interfaces, in combination with internal cognitive processes, to perceive visual input, remember information, reason using that information, and generate motor output (writing out numbers or letters). It uses these abilities to perform eight different tasks, ranging from perceptual-motor tasks (recreating the appearance of a perceived digit) to reinforcement learning (in a gambling task) to language-like inductive reasoning (completing abstract patterns in observed sequences of digits).  These tasks can be performed in any order, they are all executed by the same model, and there are no changes to the model between tasks.  To see the model perform the tasks, see \url{http://nengo.ca/build-a-brain/spaunvideos}.

We have compared the performance of Spaun to human and animal data at several levels of detail \cite{Eliasmith2012b}.  Along many metrics the two align; for example, the model and the brain share (1) dynamics of firing rate changes in striatum during the gambling task, (2) error rates as a function of position when reporting digits in a memorized list, (3) coefficient of variation of inter-spike intervals, (4) reaction time mean and variance as a function of sequence length in a counting task, (5) accuracy rates of recognizing unfamiliar handwritten digits, and (6) success rates when solving induction tasks similar to those found on the Raven's Progressive Matrices (a standard test of human intelligence), among other measures.  It is these comparisons, demonstrating the range and quality of matches between the model and real neural systems, that makes it plausible to suggest that Spaun is capturing some central aspects of neural organization and function.

The Spaun model is simulated in the neural simulation package Nengo \cite{Bekolay2014}, which natively implements the NEF and SPA methods.

\subsection{The compartmental model}

\begin{figure}
  \centering
  \includegraphics[width=\columnwidth]{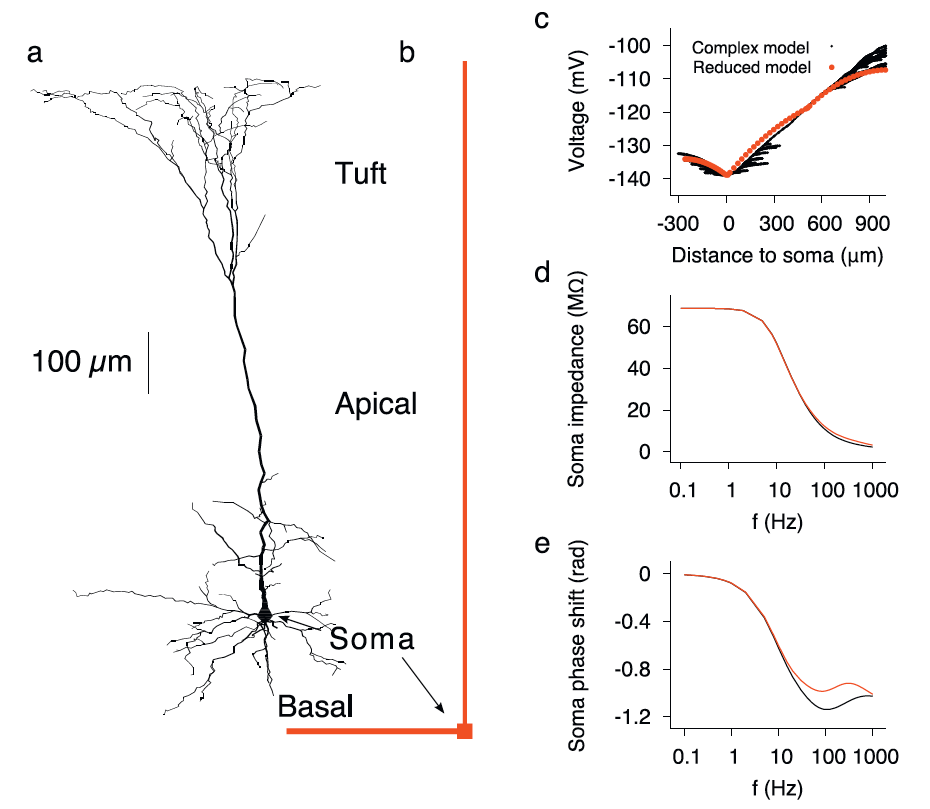}
  \caption{The reduced model compared to the complex model in terms of passive response. a) The morphology of a neuron from \cite{Stuart1998} used to create the reduced model. b) An illustration of the functional sections, defined over 20 compartments, used to optimize the reduced model. c) Voltage induced at different locations in the models as a function of a -1nA current injected in the soma. d) The somatic impedance as a function of low-amplitude oscillatory somatic input current for both models. (e) The soma phase-shift between the oscillatory input current and  membrane potential oscillation for both models. Reproduced from \cite{Bahl2012}.}
  \label{fig-neuron}
\end{figure}

To replace the simple LIF neurons in our model, we have chosen a compartmental conductance model of cortical neurons that is similar in complexity to those used in the recent HBP model \cite{Markram2015}. In the HBP model, neurons had up to 13 ion channel types and 4 compartments. In this work, we use the pyramidal cell model developed and described in detail in \cite{Bahl2012}.  This model is  reduced from a model with several hundred compartments, based on the neural reconstructions in \cite{Stuart1998}.  The reduced model that we are using has 20 compartments across 4 functional areas (soma, basal dendrite, apical dendrite, and apical dendritic tuft), 27 parameters, and 9 different ion channels.  After automatically choosing parameter values based on an optimization method, the reduced model very closely replicated the behavior of the complex model (see, e.g., Figure \ref{fig-neuron}). The model is available for download at \url{http://senselab.med.yale.edu/ModelDB}, and runs in the NEURON 7.1 simulator.

To replace neurons in the Spaun model with this compartmental model, we 
developed a Python interface, called \emph{nengo\_detailed\_neurons}, between 
Nengo and NEURON.  Consequently, all of BioSpaun except for the compartmental 
neurons were run in Nengo, while the Python interface to NEURON was used to 
communicate between the two simulators.  The Python interface is available at 
\url{https://github.com/nengo/nengo_detailed_neurons}.

The NEF is defined such that the response properties of the specific neuron 
models being employed can be taken into account during the optimization process 
for determining connection weights.  Typically, point neurons are used so all 
connections are to the same compartment.  In the case of employing compartmental 
models, with different compartments for different areas of the cell, the weights 
of excitatory synapses were uniformly distributed along the compartment modeling 
the apical dendrite.  The synaptic weights were linearly scaled with the 
distance from the soma (up to a factor of 2 if at the end of the dendritic 
compartment).  This method approximately preserved the somatic effects of arriving 
spikes, resulting in neuronal activity that can be accounted for using the 
standard NEF methods.  We intend to replace this simple heuristic with more 
precise methods in the future. The inhibitory synapses directly 
targeted the somatic compartment.

We exploited the inclusion of this more sophisticated neuron model by testing the effects of TTX on high-level function.  TTX is known to block voltage-gated sodium channels in neurons.  Consequently, to simulate the effects of TTX on the compartmental neurons, different percentages of the sodium channels were blocked to simulate different concentrations of the drug.

\section{Results}
\subsection{Comparison of LIF and compartmental information processing}

\begin{figure}
  \centering
  \includegraphics[width=\columnwidth]{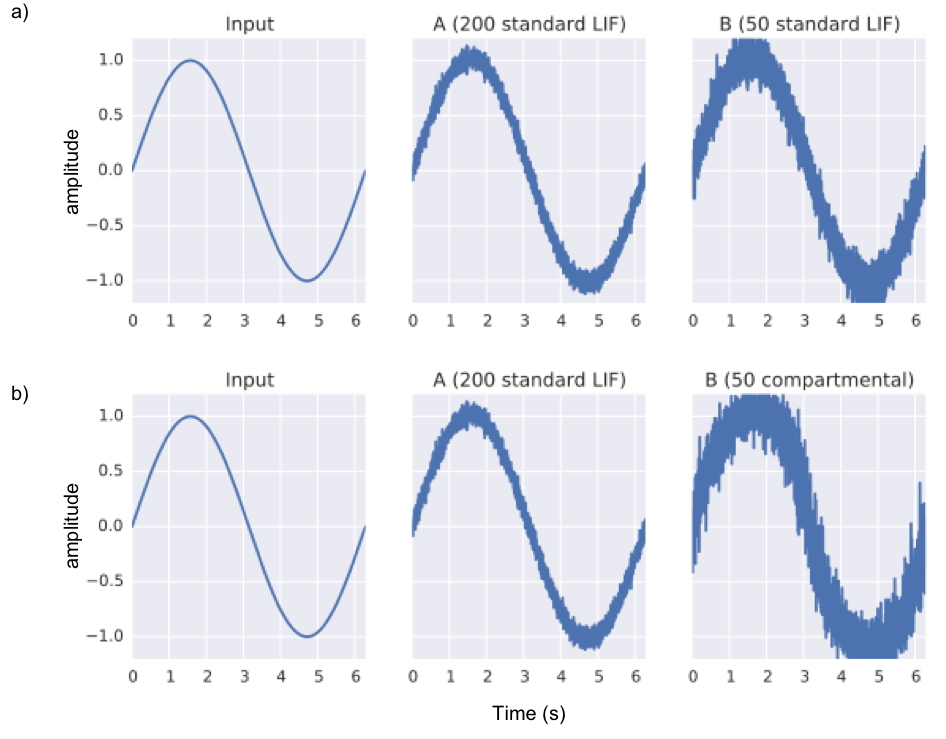}
  \caption{Simple information processing through a communication channel in LIF and compartmental neurons.  The first panel is the state variable input, which is encoded into spikes by population A. The second panel shows the estimate of the state variable from decoding spikes from population A. The third panel shows the estimate of the state variable from decoding the spikes from population B. a) The channel implemented with LIF neurons in both populations using standard NEF methods. b) The same channel implemented with compartmental neurons in the output population.}
  \label{fig-sinewave}
\end{figure}

To ensure that the NEF method was successfully applied to populations of compartmental neurons, we constructed a circuit consisting of 200 input neurons in population A and 50 output neurons in population B.  The function computed by the circuit is identity, implementing a communication channel.  We injected one period of a sine wave into the input population and reconstructed the input from both LIF and compartmental neurons (see Figure \ref{fig-sinewave}).

\begin{figure}
  \centering
  \includegraphics[width=\columnwidth]{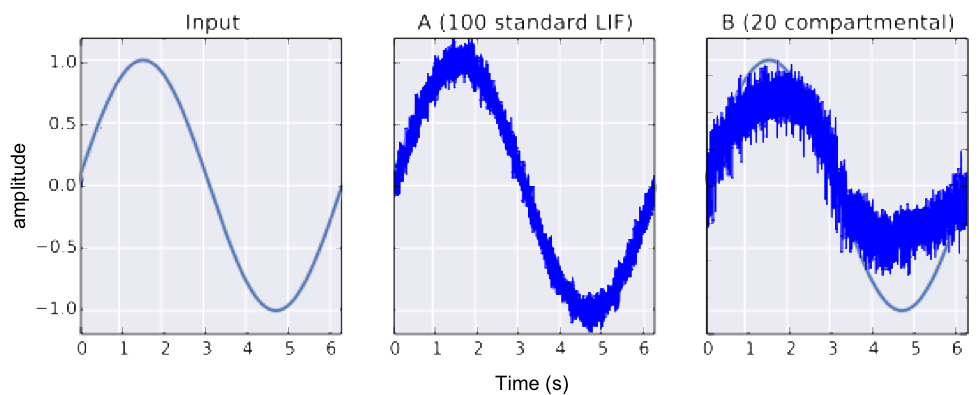}
  \caption{TTX application to a communication channel. The same circuit as shown in Figure \ref{fig-sinewave}b with 60\% of the sodium channels blocked.}
  \label{fig-simplettx}
\end{figure}

Both circuits are able to represent the input well, although the compartmental neurons are slightly noisier.  Specifically, the RMS error in the LIF circuit was 0.1 whereas the RMS error in the compartmental circuit was 0.21.  To demonstrate the effects of a high dose TTX application, we blocked 60\% of the sodium channels in the conductance neurons in this circuit (Figure \ref{fig-simplettx}.  This manipulation significantly negatively impacts the response of the neurons and increases the RMS error.

\subsection{Effects of TTX application on visual processing}

In the original Spaun model, the visual system consists of a ventral stream model that includes V1, V2, V4 and IT.  In the digit recognition task, Spaun classifies human handwritten digits from the MNIST database based on the neural representation in area IT.  To initially examine the effects of using the compartmental neurons in Spaun, we replaced the LIF neurons in IT with compartmental neurons.  This is a much more challenging task than a simple communication channel because classification is a highly nonlinear function.  A video of normal performance on the digit classification task can be found at \url{https://youtu.be/QDSXhuPGHSs}.

\begin{figure}
  \centering
  \includegraphics[width=.7\columnwidth]{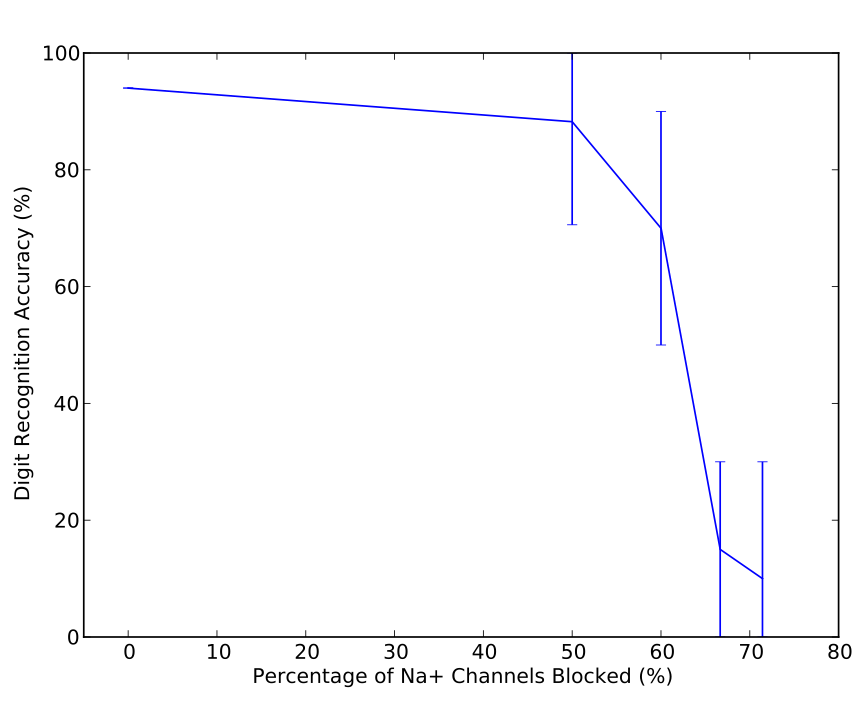}
  \caption{Handwritten digit recognition accuracy as a function of TTX dosage in BioSpaun. Error bars are 95\% confidence intervals.}
  \label{fig-digits}
\end{figure}

The original classification accuracy of Spaun (94\%) was preserved with the introduction of the compartmental models (see Figure \ref{fig-digits}, leftmost value). We examined the effects of changing the dosage of TTX applied to these model neurons on classification accuracy.  The results are shown in Figure \ref{fig-digits}. As can be seen, classification is robust to large numbers of blocked sodium channels, with little decrease in accuracy at 50\% channel blockage.  However, by 72\% blockage, the model's accuracy is at chance levels (i.e. 10\%).

\subsection{Effects of TTX applied to frontal cortex across tasks}

Application of TTX to visual cortex has a similar consequence across all of the tasks performed by Spaun: the input is made noisier and less certain.  This is unsurprising given the kind of effects demonstrated in the previous section.  However, those effects are demonstrated in a largely feedforward network.  As well, the effects are only explored in a single task.  Here we apply the manipulation to a recurrent network and examine the effects the same TTX manipulation across different tasks.

\begin{figure}
  \centering\includegraphics[width=\columnwidth]{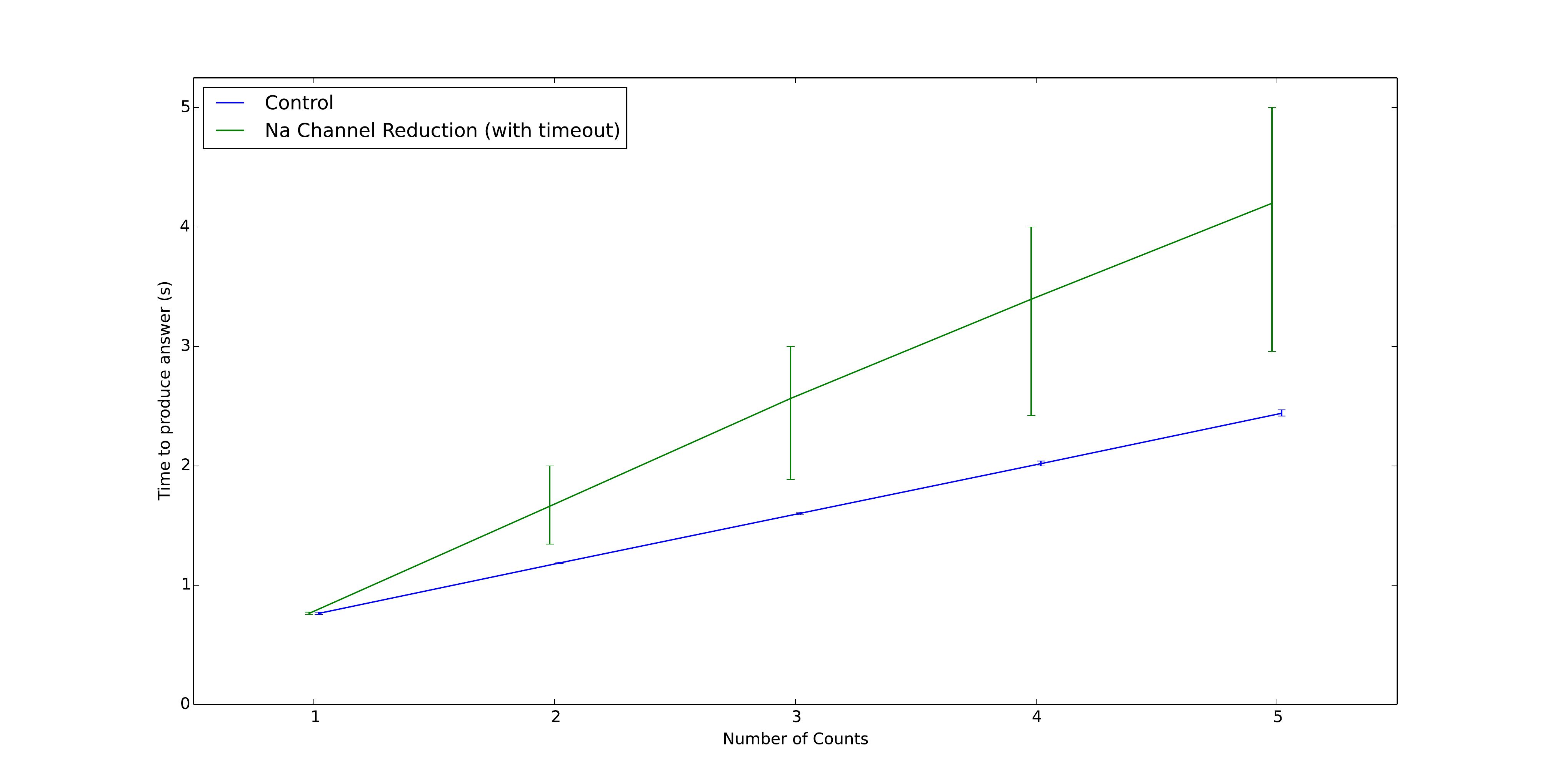}
  \caption{Effect of TTX application on the counting task time to completion. The control task reproduces the results of \cite{Eliasmith2012b} where it was shown to statistically match human performance times, while using detailed neurons.  The effetcs of TTX application is shown by the green line, which shows significantly worse timing, with many time outs.  Error bars are 95\% confidence intervals.}
  \label{fig-count}
\end{figure}

\begin{figure}
  \centering\includegraphics[width=\columnwidth]{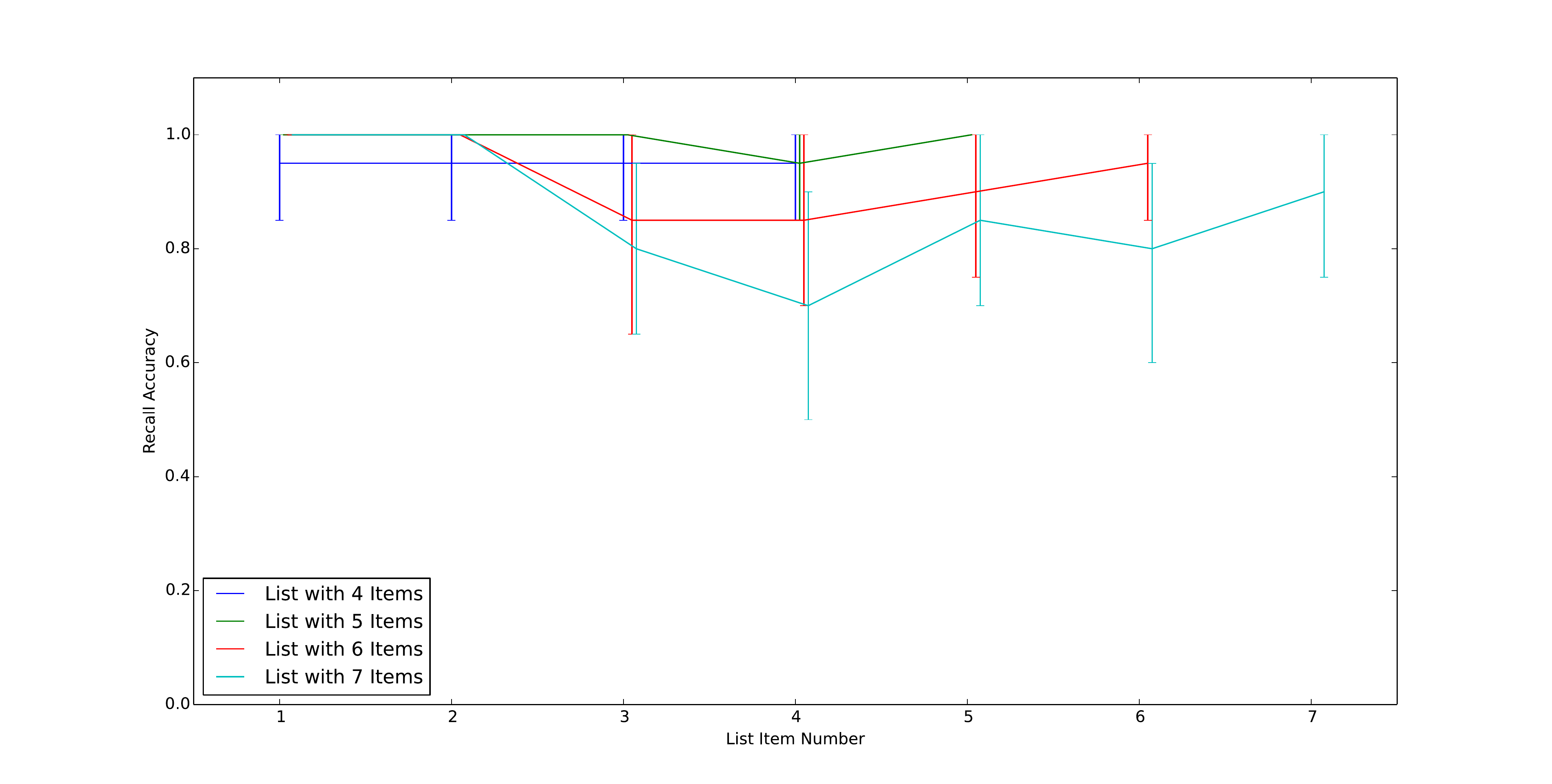}
  \centering\includegraphics[width=\columnwidth]{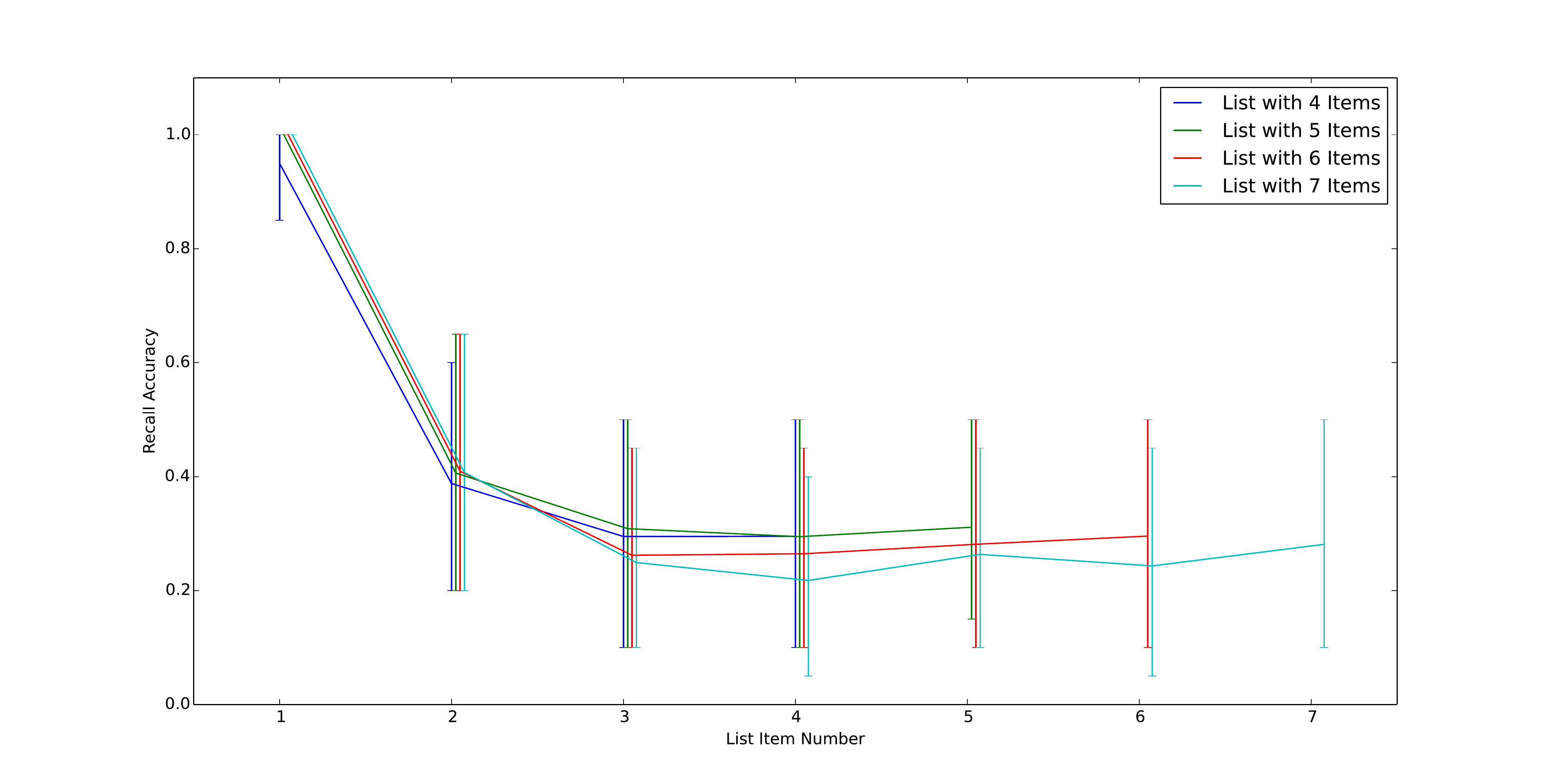} 
  \caption{Effect of TTX application on the list memory task. Top: Performance with no TTX application reproduces the results reported in \cite{Eliasmith2012b}, which are statistically indistinguishable from human performance, while using detailed neurons.  Bottom:  The effects of the same TTX application as in Figure \ref{fig-count}.  In this case the recency effect is largely destroyed, and after two items performance is at chance.  Error bars are 95\% confidence intervals.}
  \label{fig-wm}
\end{figure}

To examine heterogeneous effects of the same manipulation in a recurrent network, we introduced conductance neurons into part of the frontal cortex of Spaun.  Specifically, the part of frontal cortex that we changed is mapped to OFC and acts as a memory that is responsible for keeping track of the task currently being performed.  As before, this introduction made no observable functional difference across tasks (see Figures \ref{fig-count} and \ref{fig-wm}).

We subsequently introduced TTX to block about 20\% of the sodium channels in this frontal area.  We examined the performance on two cognitive tasks, the counting task and the list memory task.  The counting tasks consists of showing BioSpaun a starting digit (e.g. 3) and a count digit (e.g. 4), and then asking for the result of counting by the count digit from the starting digit (e.g. 7).  Spaun replicated human reaction time data on this task.  An example run of the performance of BioSpaun on the counting task is shown in this video \url{https://youtu.be/FoOGqzG8_WU}.  As shown, BioSpaun is able to encode the starting and count digits, and begins the internal counting process, but that process is interrupted (i.e. the task state is forgotten) before the counting completes.  Consequently, BioSpaun produces no result.  Figure \ref{fig-count} demonstrates performance across various numbers of counts.  These results demonstrate that the performance significantly worsens with the application of TTX, both in terms of reaction time, and task completion.  Notably, there was a timeout of 2s to complete the task after the display of the last digit.  Consequently, the worst possible performance (i.e., not responding to the task) was most common for 5 counts with TTX.

The second task we performed the same manipulation in is the list memory task. In the list memory task, BioSpaun is shown a list of digits (with varying numbers of digits), which it must reproduce after a delay.  The original Spaun model captures primacy and recency effects in such a task. Performance of BioSpaun under the influence of TTX is shown in this video \url{https://youtu.be/kpwoBccdmd8}.  In this case, BioSpaun is able to encode the list correctly, and begins to write out digits but is interrupted before completing the list.  Performance across lists of various lengths is shown in Figure \ref{fig-wm}.

Although the behavior across these tasks is in someways quite different, as the model always produces some output in the list task, while often not at all in the counting task, it is clear how these different behaviors are the result of the same TTX-induced failure.  That is, TTX consistently induces a failure to retain the task state long enough to complete the given task.  While preliminary, this is a useful demonstration of how a single brain model can explain the behaviorally different effects of a single underlying molecular manipulation.

\section{Conclusion}
\label{S:conclusion}

Many reasons have been offered as to why large-scale models are important to build. These include the ability to understand mysterious brain disorders, from autism to addiction \cite{just2012}, to develop and test new kinds of medical interventions, be they drugs or stimulation \cite{beeler2012}, and to provide a way to organize and unify the massive amounts of data generated by the neurosciences \cite{tyrcha2013}. However, no question about brains seems to loom larger than: “How do brains control behavior?” The vast majority of sophisticated behavior is the result of the interactions between many brain areas, recruiting many millions of cells, each of which exhibits complex nonlinear dynamics. Without constructing models that explore these complex interactions, and how they relate to those dynamics, we are unlikely to be able to understand how to help a distressed brain, or explain how biological mechanisms give rise to cognitive behavior. In short, without large-scale models, we cannot test large-scale hypotheses.

Until the Spaun model, past work on large-scale models has been surprisingly silent on the connection between complex neural activity and observable behavior.  Even with the introduction of Spaun, the link between low-level biophysical properties and cognitive behavior was only mildly elucidated because of the simplicity of the single cell models employed.  With BioSpaun, we have provided a preliminary but suggestive method for simulating brain models that include extensive biophysical detail while not sacrificing a clear connection to interesting behavior.  We have only exploited this connection to examine potential consequences of employing a drug whose effects are reasonably approximated through a simply mechanism (i.e. local sodium channel blockage).  Nevertheless, we believe that this kind of model opens the possibility of examining how a wide variety of low-level biological interventions can influence systemic behavior.


\section{References}
\bibliographystyle{model1-num-names}
\bibliography{biospaun.bib}

\end{document}